\documentclass[twocolumn,letterpaper]{igs}

\usepackage{igsnatbib}
\usepackage{stfloats}
\usepackage{color}
\usepackage[intlimits]{amsmath}
\usepackage{amssymb}
\usepackage{calc}
\usepackage{float}
\usepackage{prettyref}
\usepackage[pdftex,breaklinks=true,colorlinks=true,linkcolor=blue,citecolor=blue,urlcolor=blue,
filecolor=blue,pdffitwindow,backref=true,pagebackref=false,bookmarks=false,bookmarksopen=true,
bookmarksnumbered=true]{hyperref}

\ifx\pdftexversion\undefined
\usepackage[dvips]{graphicx}
\else
\usepackage[pdftex]{graphicx}
\fi

\usepackage{soul}  

\newrefformat{fig}{\hyperref[#1]{Fig.\,\ref*{#1}}}
\newrefformat{sec}{\hyperref[#1]{Sec.\,\ref*{#1}}}
\newrefformat{tab}{\hyperref[#1]{Tab.\,\ref*{#1}}}
\newrefformat{app}{\hyperref[#1]{App.\,\ref*{#1}}}
\newrefformat{subsec}{\hyperref[#1]{Sec.\,\ref*{#1}}}
\newrefformat{eq}{\hyperref[#1]{(\ref*{#1})}}

\def\figpath{./figs} 

\newcommand{\CV}{\text{CV}}
\newcommand{\EW}[2][]{\text{E}_{#1}\left[#2\right]}
\newcommand{\ex}{\text{ex}}
\newcommand{\jour}{\text{d}}
\newcommand{\km}{\text{km}}
\newcommand{\logt}[1]{\text{log}_{10}\left(#1\right)}
\newcommand{\Max}{\text{max}}
\newcommand{\meter}{\text{m}}
\newcommand{\Min}{\text{min}}

\newcommand{\percent}{\text{\%}}
\newcommand{\pl}{\text{pl}}
\newcommand{\seconds}{\text{s}}
\newcommand{\SD}{\text{SD}}

\restylefloat{figure}
\def\floatpos{!h}

\newcommand{\singlecolumnfigure}[3]{
  \begin{figure*}[\floatpos]%
    \centering
    \includegraphics[width=8.5cm]{#1}
    \caption{#2}
    \label{#3}
  \end{figure*}
}
\newcommand{\doublecolumnfigure}[3]{
  \begin{figure*}[\floatpos]
    \centering
    \includegraphics[width=17.8cm]{#1}
    \caption{#2}
    \label{#3}
  \end{figure*}
}

\newcommand{\singlecolumntable}[3]{
  \begin{table*}[\floatpos]
    \centering
    #1
    \caption{#2}
    \label{#3}
  \end{table*}
}

\begin{document}

\title{%
  The variability of tidewater-glacier calving:\\ 
  origin of event-size and interval distributions
}

\shorttitle{Variability of tidewater-glacier calving} 

\author[Chapuis \& Tetzlaff]{%
  Anne CHAPUIS$^1$, 
  Tom TETZLAFF$^{1,2,\#}$
}

\affiliation{%
$^{1}$Dept.~of Mathematical Sciences and Technology, Norwegian University of Life Sciences, {\AA}s, Norway\\
$^{2}$Inst.~of Neuroscience and Medicine (INM-6), 
Computational and Systems Neuroscience \& 
Inst.~for Advanced Simulation (IAS-6), Theoretical Neuroscience, 
J\"ulich Research Centre and JARA, J\"ulich, Germany\\
$^{\#}$ Corresponding author: 
{\normalfont\url{t.tetzlaff@fz-juelich.de}}
}

\abstract{%
  Calving activity at the termini of tidewater glaciers 
  produces a wide range of iceberg sizes at irregular intervals.
  We present calving-event data obtained from continuous observations
  of the termini of two tidewater glaciers on Svalbard, and show that
  the distributions of event sizes and inter-event intervals can be
  reproduced by a simple calving model focusing on the mutual
  interplay between calving and the destabilization of the glacier
  terminus. The event-size distributions of both the field and the model
  data extend over several orders of magnitude and resemble power
  laws.  The distributions of inter-event intervals are broad, but
  have a less pronounced tail. In the model, the width of the size
  distribution increases with the calving susceptibility of the
  glacier terminus, a parameter measuring the effect of calving on the
  stress in the local neighborhood of the calving region.  Inter-event
  interval distributions, in contrast, are insensitive to the calving
  susceptibility. Above a critical susceptibility, small perturbations
  of the glacier result in ongoing self-sustained calving activity.
  The model suggests that the shape of the event-size distribution of
  a glacier is informative about its proximity to this transition
  point. Observations of rapid glacier retreats can be explained by
  supercritical self-sustained calving.
}

\maketitle

\noindent
\section{Introduction}
\label{sec:intro}

Iceberg calving plays a key role in glacier dynamics, and hence in
how tidewater glaciers and ice sheets respond to climate change,
thereby impacting predictions of sea level rise in the
future \citep{vanderveen97, ONeel2003, Benn2007a, Nick2009}. So far,
the mechanisms underlying the calving dynamics are only partly
understood.  To summarize the potential controls affecting iceberg
calving, \citet{Benn2007a} proposed the following classification:
\emph{first-order controls} determining the position of the glacier
terminus, \emph{second-order controls} responsible for the calving of
individual icebergs, and \emph{third-order controls} related to the
calving of submarine icebergs. The first-order control on calving is
the strain rate resulting from spatial variations in the glacier
velocity, responsible for the opening of crevasses. Crevasse formation
is reinforced by the presence of liquid water, either from surface
melt or rain events. Second-order controls are processes weakening the
glacier terminus and favoring fractures, like the presence of force
imbalances at the glacier terminus resulting from the margin geometry,
undercutting of ice and torque due to buoyancy. Third-order controls
trigger submarine iceberg calving by processes like the formation of
basal crevasses, tides and buoyancy.

\par
The majority of previous studies describes calving by means of
macroscopic variables such as the overall calving rate (calving speed),
i.e.~the total ice loss at the glacier terminus within rather long time
intervals. Corresponding models (``calving laws'') relate the dynamics
of the calving speed to parameters like the water depth
\citep{Brown82, Oerlemans2011}, the height-above-buoyancy
\citep{Vanderveen96, Vieli2001, Vieli2002}, the penetration of surface
and basal crevasses arising from the longitudinal strain near the
calving terminus and surface melt \citep{Benn2007b, Nick2010, Otero2010},
and more general glacier characteristics such as the ice thickness, the
thickness gradient, the strain rate, the mass-balance rate, and
backward melting of the terminus \citep{Amundson2010}.
\par
So far, only few studies have been dedicated to a description of
calving dynamics at the level of individual calving events.
Continuous monitoring of individual events directly at the glacier
terminus is challenging; consequently, data are sparse. Previous
studies of the calving-event statistics were based on occasional or
discontinuous observations of calving events \citep{Washburn36,
  Warren95, ONeel2003},
or on indirect measurements, e.g.~event sizes obtained from icebergs
floating in the sea \citep{Budd80,Orheim85,Wadhams88} or from seismic
activity \citep{ONeel2010}.
The available data indicate that event sizes are highly variable and
broadly distributed \citep[e.g.][]{Bahr1995,ONeel2010}.
However, distributions of event sizes obtained from floating icebergs
in the sea are likely to be biased due to melting and disintegration
\citep{Neshyba80,Bahr1995}. 
Seismic measurements can only detect large calving events reliably
\citep{Koehler12_393}. In addition, estimating calving-event sizes
(ice volume) from seismic-event magnitudes is problematic unless the
relation between these two quantities is clearly established
\citep[e.g.~through calibration by means of direct visual observations;
see][]{Koehler12_393}.
Hence, the resulting size distributions may be biased.
In principle, calving activity at the single-event scale could be
monitored by means of repeat photography \citep{ONeel2003}, laser
scanning, or ground-based radar \citep{Chapuis2010}. So far, however,
no such data have been published. Here, we present single-event data
obtained from continuous visual observations directly at the termini
of two tidewater glaciers on Svalbard. Our data confirm that both the
sizes of individual calving events and the time intervals between
consecutive events are broadly distributed. 
\par
So far, the mechanisms underlying this calving variability remain
unknown. It is unclear whether it reflects variability in external
conditions, e.g.~temperature or tides, or whether it is generated by
the internal calving dynamics itself. Fluctuations in external
conditions can hardly be controlled in nature. Disentangling these two
potential sources of variability therefore requires a model of the
calving process.  A description of the size and timing of individual
calving events is beyond reach of the macroscopic continuum models
focusing on the overall calving rate (see above). \citet{Bahr1995},
\citet{Bassis2010} and \citet{Amundson2010} proposed models accounting
for the discreteness of calving.  In the model of
\citet{Amundson2010}, calving events are triggered when the terminus
thickness decreases to some critical value. According to this model,
the event sizes and inter-event intervals are fixed for constant model
parameters. Variability in event sizes and intervals can only result
from fluctuations (e.g.~seasonal variations) in these
parameters. \citet{Bassis2010} describes the motion of the glacier
terminus in one dimension as a stochastic process. In the study of
\citet{Bahr1995}, calving is modeled as a percolation process in a
two-dimensional lattice representing a region close to the glacier
terminus. In this model, microfractures are randomly and independently
generated according to some cracking probability. Calving events occur
whenever a section of ice is surrounded by a cluster of connected
microfractures. In this model, calving in one region of the model
glacier has no effect on the state of the rest of the glacier. Both
the model of \citet{Bahr1995} and that of \citet{Bassis2010} are
inherently stochastic.
In our study, we propose an alternative calving model focusing on the
mutual interplay between calving and the destabilization of the local
neighborhood of the calving region. Although the calving dynamics of
this model is fully deterministic, it generates broad distributions of
event sizes and inter-event interval distributions which are
consistent with the field data, even under stationary conditions.
\par
Ultimately, breaking of ice, formation of fractures (crevasses) and,
therefore, calving are consequences of internal ice stress
\citep{Benn2007a}. Several mechanisms contribute to the build-up of
stress at the glacier terminus, e.g.~glacier-velocity gradients,
buoyancy, tides, or changes in the glacier-terminus geometry due to
calving itself. The model proposed in this study describes the
interplay between internal ice stress and calving as a
positive-feedback loop: the glacier calves if the internal ice stress
exceeds a critical value. The detachment of ice leads to an increase
in stress in the neighborhood of the calving region, mainly due to a
loss buttress, but also as a consequence of a reduction in ice burden
pressure, increase in buoyancy and terminus acceleration. This
increase in ice stress destabilizes the neighborhood of the calving
region, i.e.~increases the likelihood of calving. In our model, the
calving-induced change in ice stress is captured by a parameter, the
calving susceptibility. We show that the positive-feedback loop
between calving and terminus destabilization alone is sufficient to
explain the large variability of iceberg sizes and inter-event
intervals observed in the field data. In the model, all other stress
contributors are treated as ``external stress'' or described by
parameters. Keeping these parameters constant enables us to study the
glacier dynamics under (ideal) stationary conditions.
\par
The paper is organized as follows: we first describe the acquisition
of the field data, the calving model and the data analysis
(\prettyref{sec:methods}). We then present the event-size and interval
distributions obtained from the field data and show that they are
reproduced by the calving model (\prettyref{sec:variability}). The
model predicts a large variability in the calving process even under
stationary external conditions. This is supported by the field data
showing that the shape of the size and interval distributions is not
affected by registered fluctuations in climatic conditions
(\prettyref{sec:external_parameters}). Next, we discuss a prediction
of the model which may be of significance for judging the stability
of a glacier: the model glacier exhibits a critical point at which it
enters a regime of ongoing self-sustained calving
(\prettyref{sec:self_sustained}), a regime which may be related to
observations of rapid glacier retreats. Finally, we demonstrate that
the calving model is consistent with the field data in the sense that
the size of future calving events is barely predictable from past
events (\prettyref{sec:predictability}). In the last section
(\prettyref{sec:discussion}), we summarize and discuss the
consequences of our work, embed the results into the literature and
point out limitations and possible extensions.

\section{Methods}
\label{sec:methods}

\subsection{Acquisition of field data}
\label{sec:field_data_acquisition}
{\bf Study regions.}
Calving activity was monitored at two tidewater glaciers on Svalbard:
Kronebreen and Sveabreen.
Kronebreen is a grounded, polythermal tidewater glacier, located at
$78^o53N$, $12^o30E$, approximately $14\,\km$ south-east of
Ny-{\AA}lesund, western Spitsbergen
(\prettyref{fig:aerial_picture}A). It is one of the fastest tidewater
glaciers on Svalbard with an average terminus velocity between $2.5$
and $3.5\,\meter/\jour$ during the summer months
\citep{Rolstad2009}. Calving activity was monitored over a 4- (August
26th, 2008, 19:00, to September 1st, 2008, 05:11, GMT) and a 12-day
period (August 14th, 2009, 00:00, to August 26th, 2009, 16:00, GMT).
At the end of August 2008, the terminal ice cliff had an elevation
between $5$ and $60\,\meter$ above water level
\citep{Chapuis2010}. About $90\percent$ of the glacier terminus was
visible from the observation camp, located approximately $1.5\,\km$
west of the glacier terminus (\prettyref{fig:aerial_picture}, open
white triangle).
The second glacier, Sveabreen, is a $30\,\km$ long, grounded tidewater
glacier located at $78^o33N$, $14^o20E$, terminating in the northern
part of Isfjorden (\prettyref{fig:aerial_picture}B). The observation
of Sveabreen was part of a Youth Expedition program with 45
participants, lasting from July 17, 2010, 14:40, to July 21, 2010,
15:00 (GMT). The camp was located approximately $500-700\,\meter$ from
the glacier terminus and offered an unobstructed view.
\doublecolumnfigure{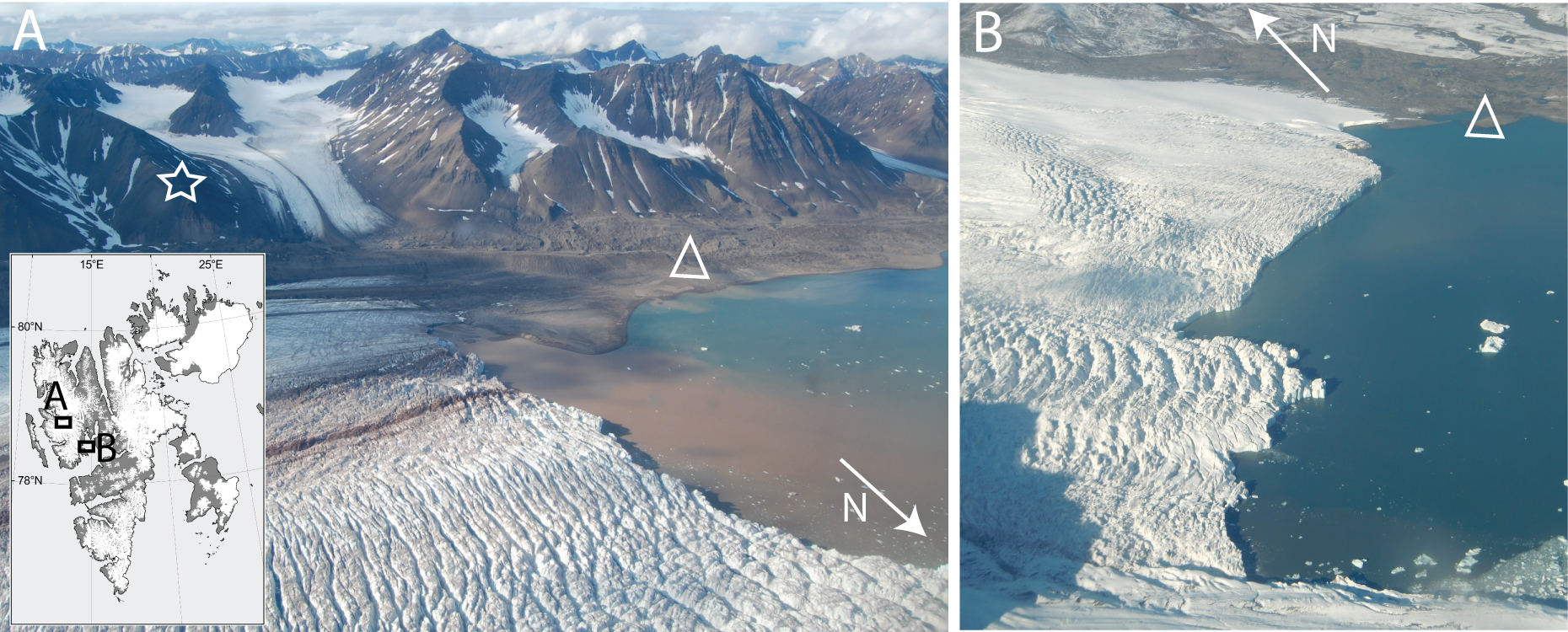}{
  Aerial pictures of Kronebreen (A) and Sveabreen (B) taken in August
  2009 and August 2010, respectively. Locations of the observation camps
  and the time-lapse camera are marked by triangles and the star,
  respectively. Inset: Map of Svalbard showing the location of the two
  glaciers.
}{fig:aerial_picture}
\par
{\bf Perceived event sizes and inter-event intervals.}
Calving activity was monitored by means of direct visual (and auditory)
inspection of the glacier termini through human observers. At both
observation sites, Kronebreen and Sveabreen, midnight sun lasts from
April 18 to August 24. Hence, calving activity could be monitored
continuously (day and night). Groups of 2--3 observers worked in
alternate shifts. Note that, despite multiple observers, each calving
event was registered only once.
For each calving event, we registered the type \citep[avalanche, block
slump, column drop, column rotation, submarine event;
see][]{ONeel2003,Chapuis2010}, location, time and perceived size. For
the data analysis, we did not distinguish between different event types.
Due to delays between the occurrence of events and the registration by
the human observers, we assign a temporal precision of $\pm{}1$
minute to the inter-event intervals $\tau$, the time between two
consecutive events.
Following the semi-quantitative approach introduced by
\citet{Warren95}, we monitored the perceived size
$\psi\in\{1,2,\ldots\}$ of each calving iceberg on an integer scale
\citep{ONeel2003}. The smallest observable events ($\psi=1$)
correspond to icebergs with a volume of about $10\,\meter^3$, the
largest ($\psi=11$) to more than $10^5\,\meter^3$ (collapse of about
$1/5$th of the glacier terminus). During common observation periods,
the perceived event sizes $\psi$ registered by different observers
could be compared. Based on these data, the error (variability) in
$\psi$ is estimated as $\pm{}1$.  
\par
{\bf Mapping perceived event sizes to iceberg volumes.}
The perceived iceberg sizes $\psi$
were mapped to the actual iceberg volumes $\mu$ by means of
photogrammetry: repeat photographs were automatically taken at
$3$-second intervals from a fixed location (star in
\prettyref{fig:aerial_picture}) using Harbotronics time-lapse cameras
\citep[see][and references therein]{Chapuis2010}. In the resulting data, we identified
$18$ calving events which were simultaneously registered by human
observers. The approximate volume $\mu$ of each event was obtained
from the estimated iceberg dimensions, and compared to the perceived
size $\psi$ (see \prettyref{fig:volume_measurement}). The relation
between $\mu$ and $\psi$ is well fit by a power law (dashed line in
\prettyref{fig:volume_measurement}; correlation coefficient $c=0.68$
in double-logarithmic representation):
\begin{equation}
  \label{eq:volume_size}
  \mu=12.6.\psi^{3.87}.
\end{equation}
Note that the empirical power-law model \prettyref{eq:volume_size} is
consistent with psychophysical findings \citep[``Stevens' power
law'';][]{Stevens57}. Using \prettyref{eq:volume_size}, we converted
the perceived iceberg sizes $\psi$ for \emph{all} visually monitored
events to an estimated volume $\mu$ (in units of $\meter^3$).
\singlecolumnfigure{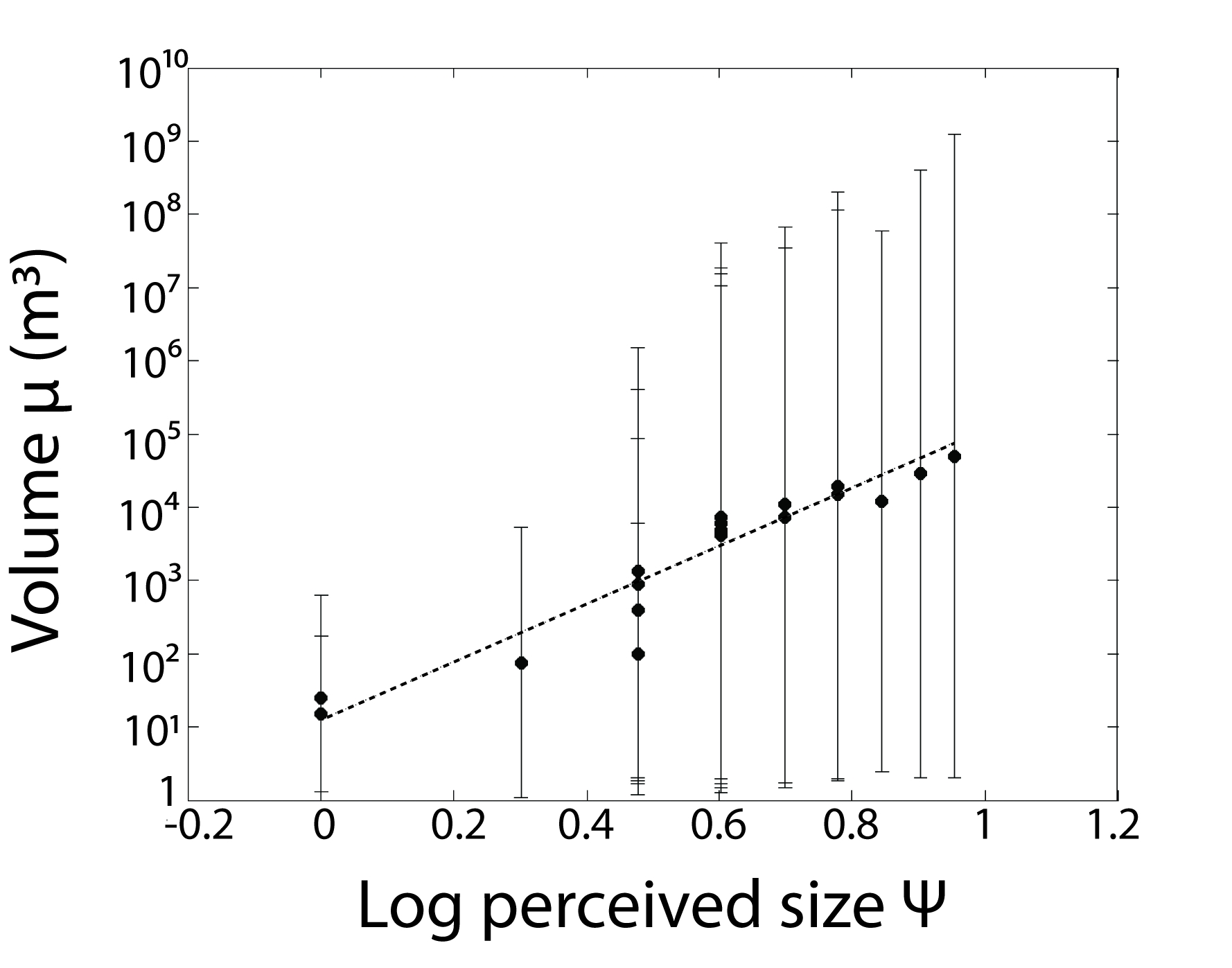}{
  Measured iceberg volume $\mu$ versus perceived iceberg size $\psi$
  (log-log scale) for $18$ individual calving events (symbols). Error
  bars depict estimated volume measurement errors. The dotted line
  represents the best power-law fit \prettyref{eq:volume_size} (linear
  fit in log-log representation).
}{fig:volume_measurement}

\subsection{Calving model}
\label{sec:calving_model} 
\par
{\bf Overview.} In our calving model, the glacier terminus is
described as a two-dimensional, discretized rectangular plane,
subdivided into cells (\prettyref{fig:model_sketch}C). Each cell
corresponds to a unit volume of ice. The state of a cell is
characterized by its internal ice stress. If this stress exceeds a
critical level, the ice breaks, the cell ``calves'' and its stress is
reset to zero (\prettyref{fig:model_sketch}A). Calving of a cell
increases the stress in neighboring cells
(\prettyref{fig:model_sketch}C) as a consequence of a loss of
buttress, a reduction in ice burden pressure, an increase in buoyancy
and terminus acceleration (see e.g.~\citealp{Benn2007a}, and
references therein). Hence, initial calving of individual cells can
trigger calving avalanches involving larger regions of the glacier
terminus. We probe the model glacier by applying small perturbations
(small stress increments) to randomly selected, individual cells. The
total number of cells participating in an avalanche triggered by a
single perturbation defines the \emph{event size} $\mu$. The time
(number of perturbations) between two consecutive events of non-zero
size corresponds to the \emph{inter-event interval} $\tau$. In the
following, the model ingredients are described in detail.
\par
{\bf Model geometry.} The calving model focuses on the calving
dynamics at the glacier terminus. For simplicity, the terminus is
described as a two-dimensional rectangular plane of width $W$ and
height $H$. The terminus is discretized, i.e.~subdivided into $WH$
cells with coordinates $\{x,y|x=1,\ldots,W;\,y=1,\ldots,H\}$ ($y=0$
and $y=H$ correspond to the sea level and the height of the glacier
terminus above sea level, respectively; see
\prettyref{fig:model_sketch}C). Each cell represents a unit volume of
ice. Note that the model neglects the third spatial dimension
perpendicular to the terminus plane.
\par 
{\bf Stress dynamics and calving.}
The internal ice stress in a cell at position $\{xy\}$ at time $t$ is
described by a scalar variable $z_{xy}(t)$
(\prettyref{fig:model_sketch}A). The cell calves at time $t^i_{xy}$ if
its internal stress exceeds a critical value of $z_\text{crit}=1$
(``yield stress''; see e.g.~\citealp{Benn2007a}, and references
therein), i.e.~if $z_{xy}(t^i_{xy})>1$ (triangle in
\prettyref{fig:model_sketch}A). The cell's calving activity can be
described mathematically as a sequence of calving times
$\{t_{xy}^i|i=1,2,\ldots\}$, or, more conveniently, as a sum of delta
pulses, \mbox{$s_{xy}(t)=\sum_i\delta(t-t_{xy}^i)$} (triangles in
\prettyref{fig:model_sketch}A,B).  After the cell at position $\{xy\}$
has calved, it is assumed to be replaced by a ``fresh'' cell
representing ice in a deeper layer. In the model, this replacement is
implemented by instantaneously resetting the stress at position $\{xy\}$
to zero (triangle in \prettyref{fig:model_sketch}A). Note that the
geometry of the model (see above) is not altered by calving.
We assume that the dynamics of the internal ice stress $z_{xy}(t)$
represents a jump process which is driven by calving of neighboring
cells and external perturbations (triangles in
\prettyref{fig:model_sketch}B). Mathematically, the (subthreshold, for
$z_{xy}\le{}1$) stress dynamics can be described by
\begin{equation} 
	  \label{eq:stress_dynamics} 
	    \frac{dz_{xy}(t)}{dt} 
	    =\sum_{k=1}^{W}\sum_{l=1}^{H} J_{kl}^{xy} s_{kl}(t) 
	    + J_\text{ext}s_\text{ext}^{xy}(t) 
	    - s_{xy}(t) 
	    \,. 
\end{equation} 
Here, the left-hand side denotes the change in stress at time $t$
(temporal derivative). The right-hand side (rhs) of
\prettyref{eq:stress_dynamics} describes different types of inputs to
the target cell $\{xy\}$. In the absence of these inputs (i.e.~if the
rhs is zero), the stress level $z_{xy}$ remains constant. The first
term on the rhs corresponds to the stress build-up due to calving in
neighboring cells: calving of cell $\{kl\}$ at time $t$ leads to an
instantaneous jump in $z_{xy}$ with amplitude $J_{kl}^{xy}$
(\prettyref{fig:model_sketch}A,B,C; see next paragraph). The second
term represents stress increments as a result of external
perturbations $s_\text{ext}^{xy}(t)$ with amplitude
$J_\text{ext}$. For simplicity, we assume that these external
perturbations are punctual events in time (delta pulses),
i.e.~\mbox{$s_\text{ext}^{xy}(t)=\sum_i\delta(t-t_{\text{ext},xy}^i)$}.
The last term on the rhs of \prettyref{eq:stress_dynamics} captures
the stress reset after calving of cell $\{xy\}$ (as described above)
and is treated as a negative input here.
Note that the single-cell calving model described here is identical to
the ``perfect integrate-and-fire model'' which is widely used to study
systems of pulse-coupled threshold elements like, for example,
networks of nerve cells \citep{Lapicque07,Tuckwell88} or sand piles
\citep{Bak87,Bak88}, or to investigate the dynamics of earthquakes
\citep{Herz95_1222}.
\doublecolumnfigure{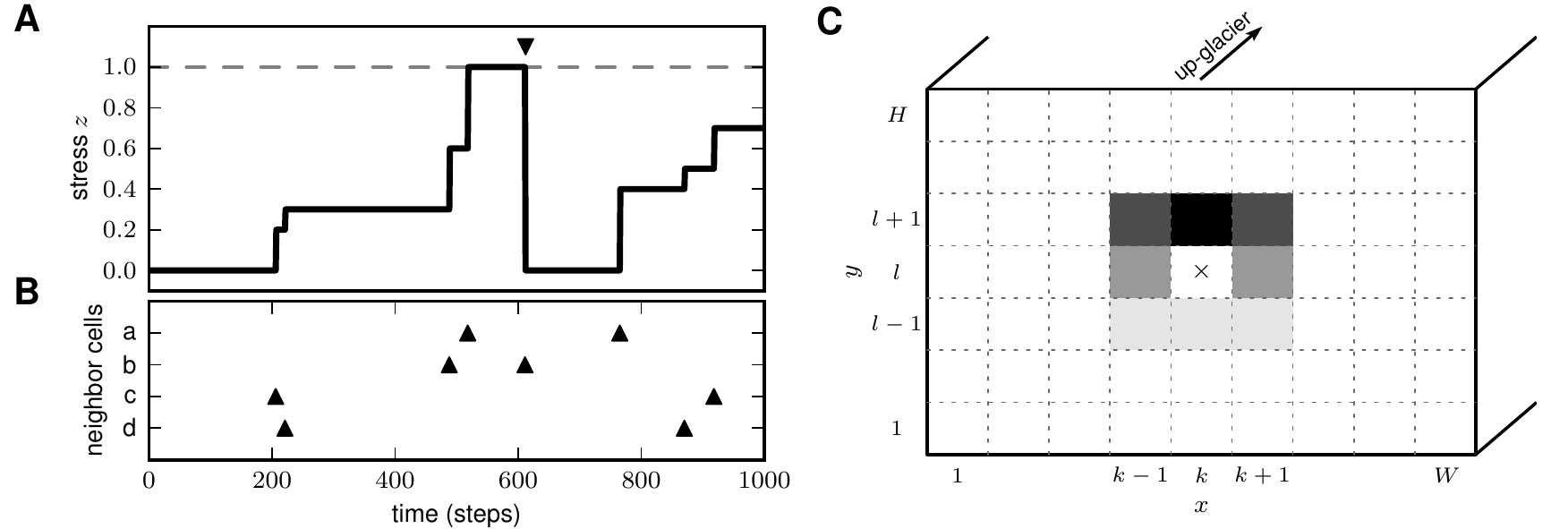}
{Sketch of the calving model. 
  A: Time evolution of internal ice stress $z$ in an individual cell. Calving of neighboring cells or 
  external perturbations (triangles shown in B) cause jumps in ice stress $z$.
  Crossing of critical stress $z=1$ (dashed horizontal line) leads 
  to calving (triangle-down marker) and reset of stress level to $z=0$.
  C: Glacier terminus (as seen from the sea/fjord; width $W$, height $H$) subdivided into $WH$ cells.
  Calving of cell $\{kl\}$ (cross) leads to stress increments (gray coded) in neighboring cells 
  (depending on relative cell position).}
{fig:model_sketch}
 \par 
 {\bf Interactions between cells.}  Calving of a cell at position
 $\{kl\}$ leads to a destabilization of its local neighborhood, mainly
 caused by a loss of buttress, but also due to local increases in
 buoyancy and changes in terminus velocity triggered by calving. In
 consequence, the stress level in neighboring cells $\{xy\}$ is
 increased (first term on rhs of \prettyref{eq:stress_dynamics}). For
 simplicity, we assume that the interaction $J_{kl}^{xy}=J(x-k,y-l)$
 between cells $\{kl\}$ and $\{xy\}$ depends only on the horizontal
 and vertical distances $p=x-k$ and $q=y-l$, respectively. Further, we
 restrict the model to excitatory (positive) nearest-neighbor
 interactions without self-coupling, i.e.
\begin{equation} 
 \label{eq:interactions} 
J(p,q)= 
 \begin{cases} 
	    0 & \text{if $p=0$ and $q=0$}\\ 
	    0 & \text{if $|p|>1$ or $|q|>1$}\\ 
	    >0 & \text{else}\,. 
	  \end{cases} 
\end{equation} 
For the results reported in the next section, we use an asymmetric 
interaction kernel (see \prettyref{fig:model_sketch}C) 
\begin{equation} 
 \label{eq:interactions_asymmetric} 
	  J(p,q)=C 
	  \begin{cases} 
	    4 & \text{if $p=0$ and $q=1$}\\ 
	    3 & \text{if $|p|=1$ and $q=1$}\\ 
	    2 & \text{if $|p|=1$ and $q=0$}\\ 
	    1 & \text{if $|p|\le{}1$ and $q=-1$}\\ 
	    0 & \text{else}\,. 
	  \end{cases} 
\end{equation} 
Here, $C$ is a normalization constant. The asymmetry in the vertical
direction reflects that cells above the calving cell will likely
experience a larger stress increment than those below due to
gravity. To test whether the dynamics of the model critically depends
on the specific choice of the interaction-kernel shape we also
consider symmetric kernels
\begin{equation} 
 \label{eq:interactions_symmetric} 
	  J(p,q)=C 
	  \begin{cases} 
	    1 & \text{if $p=0$ and $|q|=1$}\\ 
	    1 & \text{if $|p|=1$ and $q=0$}\\ 
	    0 & \text{else}\,. 
	  \end{cases} 
\end{equation} 
Qualitatively, the results for symmetric and asymmetric interaction
kernels are the same (not shown here). Note that, with the symmetric
kernel \prettyref{eq:interactions_symmetric}, our calving model is
(essentially\footnote{In the model of \citet{Bak87,Bak88}, the
  ``stress'' $z$ is reset \emph{by} a fixed amount after ``calving'',
  whereas we consider a reset \emph{to} a fixed value $z=0$.})
identical to the sandpile model of \citet{Bak87,Bak88}.
\par 
To study the dependence of the calving dynamics on the coupling
between cells, we consider the total \emph{calving susceptibility}
$w=\sum_p\sum_q J(p,q)$ as a main parameter of the model. It
characterizes the overall effect of a calving cell on the ice stress
in its local neighborhood.  An increase in the susceptibility $w$
corresponds to a \emph{destabilization} of the glacier terminus. Note
that $w$ is measured in units of the critical stress $z_\text{crit}$;
an increase in $w$ can therefore also be interpreted as a decrease in
$z_\text{crit}$. To study the effect of ice susceptibility and/or
yield stress, it is therefore sufficient to vary $w$ and keep
$z_\text{crit}=1$ constant. Both ice susceptibility and yield stress
are determined by external factors like temperature, glacier velocity,
buoyancy, glacier thickness, etc.. An increase in temperature, for
example, lowers the yield stress \citep{Benn2007a} and, thus, leads to
an increase in the susceptibility $w$.
\par
The calving susceptibility $w$ plays a key role for the dynamics of
the calving model. To illustrate this, let's assume that the states
$z_{xy}(t)$ of the cells are uniformly distributed in the interval
$[0,1]$ (across the ensemble of all cells). Calving of some cell
$\{kl\}$ at time $t$ will inevitably trigger calving in any adjacent
cell $\{xy\}$ with $z_{xy}(t)>1-J_{kl}^{xy}$. With the above
assumption, the probability of a cell $\{xy\}$ fulfilling this
condition is $J_{kl}^{xy}$. The total number of cells calving in
response to a calving cell is, on average, given by the sum
$\sum_{\{xy\}}J_{kl}^{xy}=w$ over all interaction strengths, i.e.~by
the calving susceptibility. Therefore, the calving susceptibility $w$
can be interpreted as the gain in the total number of calving cells:
If $w<1$, calving activity will quickly die out. If $w>1$, the total
number of calving cells tends to grow in time. For $w=1$, the system
is balanced in the sense that the average total number of calving
cells remains approximately constant. Strictly speaking, this holds
only under the above assumption of a uniform state distribution. As
illustrated in \prettyref{fig:model_examples}B,C, the cell states are
indeed widely distributed over the entire stress interval
$[0,1]$. Therefore, we may expect that $w=1$ marks a transition point
for the dynamics of the model. In fact, as shown in
\prettyref{sec:results}, the variability in calving-event sizes
increases substantially at $w\gtrsim{}1$ \citep[see
also][]{Shew13_88}.
\par 
{\bf Experimental protocol.} Due to the above described interactions
between cells, calving of individual cells may trigger calving in
neighboring cells, thereby causing calving avalanches.
At the beginning of each experiment, the internal ice stress of each
cell was initialized by a random number drawn from a uniform
distribution between $0$ and $1.1$. On average, $10\%$ of the cells
were therefore above the critical stress $z_\text{crit}=1$ and started
calving immediately. In general, this initial calving stopped after
some time (see, however, \prettyref{sec:self_sustained}). After this
warm-up period, we performed a sequence of perturbation experiments:
in each trial $m=1,\ldots,M$, a single cell $\{kl\}$ was randomly
chosen and perturbed by a weak delta pulse
$s_\text{ext}^{kl}(t)=\delta(t)$ of amplitude $J_\text{ext}=0.1$ (at
the beginning of each trial, time was reset to $t=0$).  The trial was
finished when the calving activity in response to the perturbation had
stopped.  We define the number of cells calving in a single trial as
the \emph{event size} $\mu$. The difference $m-u$ between the id's of
two subsequent successful trials $u$ and $m>u$ ($m,u\in[1,M]$),
i.e.~trials with $\mu>0$, defines the \emph{inter-event interval}
$\tau$. Examples of calving activity in individual trials are shown in
\prettyref{fig:model_examples}.
\par 
{\bf Model parameters.} 
The model parameters are summarized in \prettyref{tab:model_parameters}. 
\singlecolumntable{ 
  \begin{tabular}{|@{}lll|} 
    \textbf{Name} & \textbf{Description} & \textbf{Value}\\\hline 
    $W$ & width of glacier terminus & $\{100,200,400\}$ \\ 
    $H$ & height of glacier terminus & $\{25,50,100\}$ \\ 
    $z_\text{crit}$ & critical stress (yield stress) & $1$\\ 
    $w$ & calving susceptibility & $\{0.5,\ldots,1.5\}$\\ 
    $J_\text{ext}$ & perturbation amplitude & $0.1$\\ 
    $M$ & number of trials & $10000$ 
  \end{tabular} 
} 
{Model parameters. Curly brackets $\{\ldots\}$ represent parameter 
  ranges.} 
{tab:model_parameters} 
\par 
{\bf Simulation details.} 
The model dynamics was evaluated numerically using the neural-network
simulator NEST \citep[][see
\url{www.nest-initiative.org}]{Gewaltig_NEST} which has been developed
and optimized to simulate large systems of pulse-coupled
elements. Simulations were performed in discrete time
$t=0,1,2,\ldots$. Cell states were updated synchronously, i.e.~calving
activity at time $t$ increments the stress in neighboring cells at
time $t+1$.

\subsection{Data analysis}
\label{sec:statistics}
In the following, we describe the characterization of the marginal
distributions and auto-correlations of event sizes and inter-event
intervals. Field and model data were analyzed using identical
methods. Similarly, we applied identical tools to the event sizes
$\mu$ and the inter-event intervals $\tau$. Unless stated otherwise,
we will therefore not distinguish between $\mu$ and $\tau$ in this
subsection, and use $X$ as a placeholder.
\par 
{\bf Distributions of sizes and intervals.} 
The overall characteristics of the distribution of data points $X_i$
($i=1,\ldots,n$; $n=$~sample size) are given by its mean, standard
deviation $\SD$, minimum and maximum, and the coefficient of variation
\mbox{$\CV=\SD/\text{mean}$} (see \prettyref{tab:tab3}). In the case
of the inter-event intervals, the $\CV$ provides a measure of the
regularity of the calving process: while $\CV=0$ corresponds to a
perfectly regular process with delta-shaped interval distribution
(clock), $\CV=1$ is characteristic for a process with exponential
interval distribution, e.g.~a Poisson point process
\citep{Cox62}.
Histograms of the data on a logarithmic scale (relative frequency:
number of observations within an interval
$[\logt{X},\logt{X}+\logt{\Delta{}X})$ normalized by $n$) are used for
graphical illustration of the entire distributions. As shown in
\prettyref{fig:field_data}B--G, \prettyref{fig:model_data}A,D and
\prettyref{fig:grouped_field_data} (symbols), the empirical
distributions obtained this way are broad and resemble power-law or
exponential distributions. Note, however, that such histograms,
obtained by binning of finite data sets, are generally biased and
therefore not appropriate for a quantitative analysis
\citep{Clauset2009}. Here, we applied maximum-likelihood (ML) methods
\citep[see][]{Clauset2009} to quantify to what extent the field and
model data $X$ can be explained by an exponential or a power-law
distribution:
\begin{eqnarray}
  \label{eq:exp_dist}
  p_\ex(X)&=N_\ex
  \begin{cases}
    e^{-\lambda{}X}& 0\le{}X_\Min\le{}X\le{}X_\Max\\
    0 & \text{else}
  \end{cases}\\
  \label{eq:pl_dist}
  p_\pl(X)&=N_\pl
  \begin{cases}
    X^{-\gamma}& 0\le{}X_\Min\le{}X\le{}X_\Max\\
    0 & \text{else}
  \end{cases}.
\end{eqnarray}
The cutoffs were set to the observed minimum and maximum,
respectively: $X_\Min=\min_i(X_i)$, $X_\Max=\max_i(X_i)$. The
prefactors $N_\pl$ and $N_\ex$ are normalization constants. The
exponents $\lambda$ and $\gamma$ were obtained by maximizing the
log-likelihoods $l_{\ex/\pl}=\EW[i]{\log(p_{\ex/\pl}(X_i))}$ for the
two model distributions. Here,
$\EW[i]{\ldots}=\frac{1}{n}\sum_{i=1}^n\ldots$ denotes the average
across the ensemble of data points. The quality of the ML fits
(goodness-of-fit test) was evaluated as described by
\cite{Clauset2009} using surrogate data and Kolmogorov-Smirnov
statistics. The resulting $p$-values indicate how well the data can be
explained by the model distributions $p_\ex(X)$ or $p_\pl(X)$. The
log-likelihood ratio $R=l_\pl-l_\ex$, the difference between the
maximum log-likelihoods, is used to judge which of the two hypotheses,
the power-law or the exponential model, fits the data better. $R>0$
indicates that the power-law model $p_\pl(X)$ is superior (and vice
versa for $R<0$). The variance of $R$, estimated as
$\EW[i]{(R_i-R)^2}$ with $R_i=\log(p_{\pl}(X_i))-\log(p_{\ex}(X_i))$
for the best-fit distributions, was used to test whether the measured
log-likelihood ratio $R$ differs significantly from zero \citep[for
details, see][]{Clauset2009}.
\par 
{\bf Auto-correlations.} 
To investigate whether calving event sizes and intervals are
informative about future events, we calculated the normalized
auto-correlations (see \prettyref{fig:correlations})
\begin{equation}
  \label{eq:autocorrelation}
  a(i)=\frac{\EW[j]{\tilde{X}(j)\tilde{X}(j+i)}}
  {\EW[j]{\tilde{X}(j)^2}}
\end{equation}
with $\tilde{X}_j=X_j-\EW[j]{X_j}$.
\par 
{\bf Software.} 
The data analysis was performed using Python
(\url{http://www.python.org}) in combination with NumPy
(\url{http://numpy.scipy.org}) and SciPy
(\url{http://scipy.org}). Results were plotted using matplotlib
(\url{http://matplotlib.sourceforge.net}).

\section{Results}
\label{sec:results}
Calving at the termini of tidewater glaciers often occurs as a sequence
of events, thereby showing the characteristics of avalanches: an
initial detachment of small ice blocks can cascade to events of
arbitrary size (see \prettyref{fig:typical_calving_event}). In this
article, we propose that the underlying dynamics can be understood as
a result of the mutual interplay between calving and the
destabilization of the local neighborhood of the calving region.
By means of a simple calving model (see \prettyref{sec:calving_model}), we show that this mechanism is sufficient to understand the
variability in event sizes and inter-event intervals observed in the
field data (see \prettyref{sec:variability}).
Fluctuations in external parameters may additionally contribute to the
calving variability but are not required to explain the data. This is
confirmed by our observation that changes in air temperature and tides
do not affect the shape of the size and interval distributions
obtained from the field data (see
\prettyref{sec:external_parameters}).
The simple calving model enables us to study the effect of glacier
parameters on the distributions of event sizes and intervals in a
controlled manner. An increase in the calving susceptibility leads to
broader event-size distributions. At a critical susceptibility, the
model glacier undergoes a transition to a regime where a small
perturbation leads to ongoing calving activity (see
\prettyref{sec:self_sustained}).
Finally, we show that the simple calving model is consistent with the
field data in the sense that the size of future calving events is not
correlated with the size of past events. Predicting event sizes from
past events is thus difficult, if not impossible (see
\prettyref{sec:predictability}).
\doublecolumnfigure{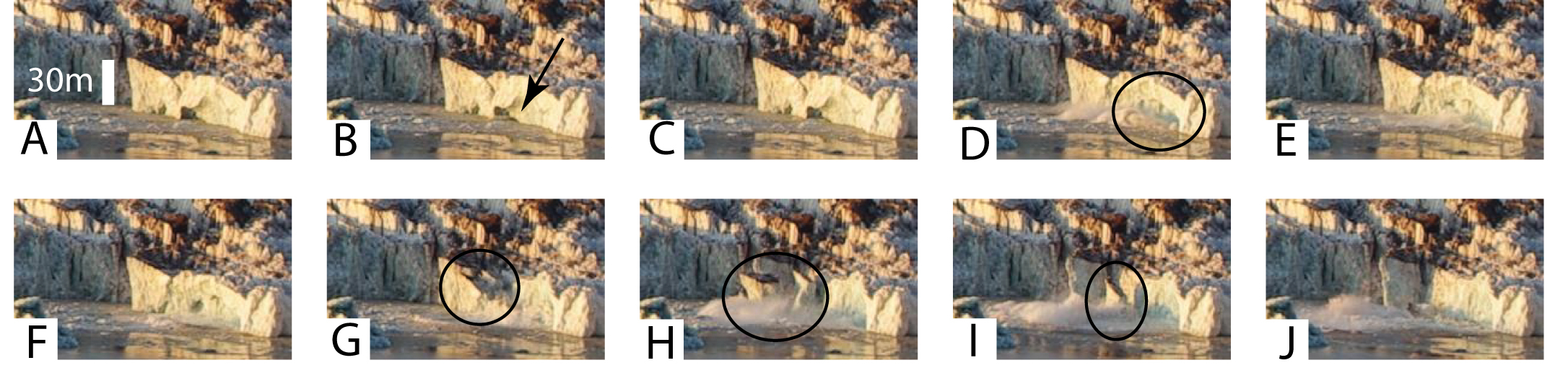}{
  Typical calving sequence observed on August 16, 2009, 21:46 GMT, at
  Kronebreen.
  The detachment of small ice blocks (B,C) triggers a large column
  drop with the entire height of the terminus collapsing vertically
  (D--G), followed by large column-rotation events with blocks of ice
  rotating during their fall (H--J).
  Time distance between consecutive images is $3\,\seconds$. Black
  arrow and ellipses mark location of individual calving events.
} {fig:typical_calving_event}

\subsection{Variability of event sizes and inter-event intervals}
\label{sec:variability}
We analyzed data obtained from two glaciers on Svalbard, Kronebreen
and Sveabreen, during three continuous observation periods with more
than $7000$ calving events in total. The longest observation period
lasted $12$ days with $5868$ events (\prettyref{fig:field_data}A). An
overview of the three data sets and the basic event-size and interval
statistics is provided in \prettyref{tab:tab3}.
\par
The sizes $\mu$ of monitored events are highly variable. They extend
over $4$ orders of magnitudes, from about $10\,\meter^3$ up to more
than $10^5\,\meter^3$ (symbols in \prettyref{fig:field_data}B--D). The
event-size coefficient of variation ($\CV$) varies between $6.2$ and
$8.2$; the standard deviations are substantially larger than the mean
values ($542$--$1512\,\meter^3$). The distributions of event sizes
exhibit long tails and resemble power laws
\prettyref{eq:pl_dist}. Maximum-likelihood (ML) fitting yields
power-law exponents $\gamma_\mu$ between $1.7$ and $2$ (solid gray
lines in \prettyref{fig:field_data}B--D). Note, however, that the
$p$-values of the goodness-of-fit test (see
\prettyref{sec:statistics}) are in all cases very small, thereby
indicating that the power-law model does not perfectly explain the
event sizes.  Still, the power-law model \prettyref{eq:pl_dist} fits
the event sizes $\mu$ better than the exponential model
\prettyref{eq:exp_dist} (dashed curves in
\prettyref{fig:field_data}B--D); the log-likelihood ratios $R$ are in
all cases significantly greater than zero ($R=2.3$ to $14$).
\par
The inter-event intervals $\tau$ span more than $2$ orders of
magnitude ($1\,\min$ up to more than $400\,\min$; symbols in
\prettyref{fig:field_data}E--G). The standard deviations exceed the
mean durations between two events ($3$--$17\,\min$) by a factor of
$\CV=1.5$ to $1.9$. Hence, the calving process is highly irregular;
substantially more irregular than a Poisson point process with
exponential interval distribution \citep[$\CV=1$;][]{Cox62}. The
interval distributions have a longer tail than predicted by the
exponential model \prettyref{eq:exp_dist} (dashed curves in
\prettyref{fig:field_data}E--G), but fall off more rapidly than the
power-law model \prettyref{eq:pl_dist} (solid gray lines in
\prettyref{fig:field_data}E--G). The log-likelihood ratios
$R\approx{}0$ confirm that the inter-event intervals $\tau$ are
neither exponentially nor power-law distributed.
\singlecolumntable{
\begin{tabular}{|@{}llll|}
\hline
 &     \multicolumn{2}{l}{{\bf Kronebreen}} & {\bf Sveabreen}\\
                                       &{\bf 2008}    &{\bf 2009}    &{\bf 2010}   \\ \hline
Total number of events	               &1041          &5868          &386     \\
Observation duration (days)            &4	      &12	     &4       \\
Start date	                       &26 Aug.       &14 Aug.       &17 July \\
End date	                       &1 Sept.       &26 Aug.       &21 July \\
\\
{\bf Event sizes $\mu$}\\			
Mean ($\meter^3$)                      &803  	      &542	     &1512   \\
Standard deviation $\SD$ ($\meter^3$)  &6599	      &4014	     &9363   \\
Coefficient of variation $\CV$         &8.2	      &7.4	     &6.2   \\  
Minimum	($\meter^3$)                   &13	      &13	     &13     \\
Maximum	($\meter^3$)                   &135070	      &135070	     &135070 \\
\\
{\bf Inter-event intervals $\tau$}\\			
Mean (min) 	                       &8	      &3	     &17  \\
Standard deviation $\SD$ (min)         &14	      &5	     &32  \\
Coefficient of variation $\CV$         &1.7	      &1.5	     &1.9 \\  
Minimum	(min)                          &1	      &1	     &1   \\
Maximum	(min)                          &201	      &98	     &446 \\
\hline
\end{tabular}
}
{Overview of the three field-data sets with event-size and interval statistics.}
{tab:tab3}
\doublecolumnfigure{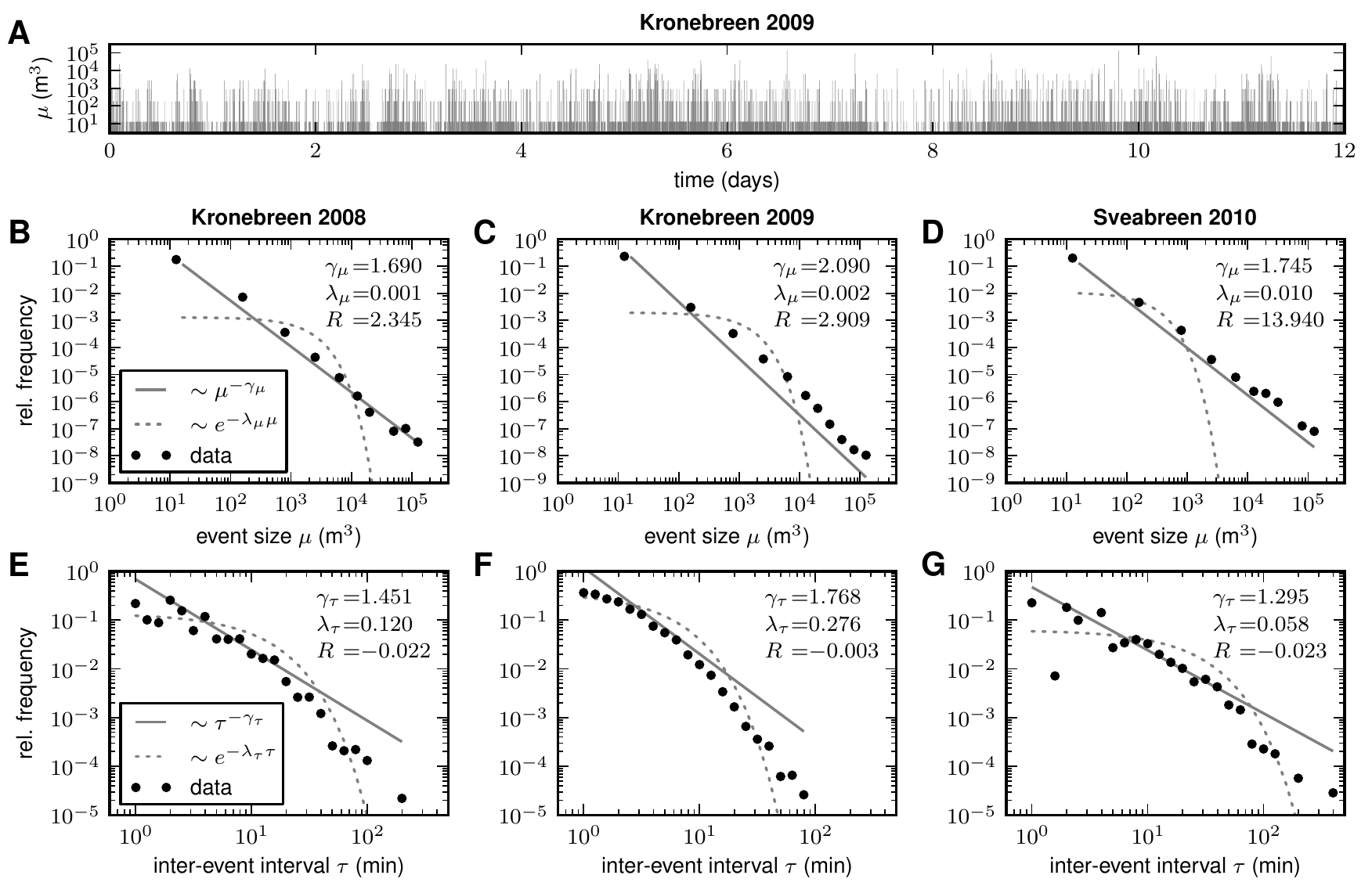}  
{
  Field data.
  A: Example time series for 12 observation days at Kronebreen (2009). 
 Each bar represents a calving event of size $\mu$ (log scale).
  B--G: Distributions (log-log scale) of iceberg sizes $\mu$ (B--D) and inter-event intervals $\tau$ 
  (E--G) for Kronebreen 2008 (B,E), 2009 (C,F) and Sveabreen 2010 (D,G). 
  Field data (symbols), best-fit power-law (solid lines; decay exponents $\gamma_\mu$, $\gamma_\tau$) 
  and exponential distributions (dashed curves; decay exponents $\lambda_\mu$, $\lambda_\tau$).
  $R$ represents corresponding log-likelihood ratio.
}
{fig:field_data}
\par
The simple calving model reproduces well the characteristics of the
event-size and interval distributions obtained from the field data. An
example time series generated by the calving model is depicted in
\prettyref{fig:model_examples}A for a susceptibility of $w=1.3$. Small
random perturbations of the model glacier lead to responses with
broadly distributed magnitudes. In most cases, there is no response at
all ($\mu=0$). Small events are frequently triggered (see example in
\prettyref{fig:model_examples}B,D,F), whereas large events are rare (see
example in \prettyref{fig:model_examples}C,E,G).
\doublecolumnfigure{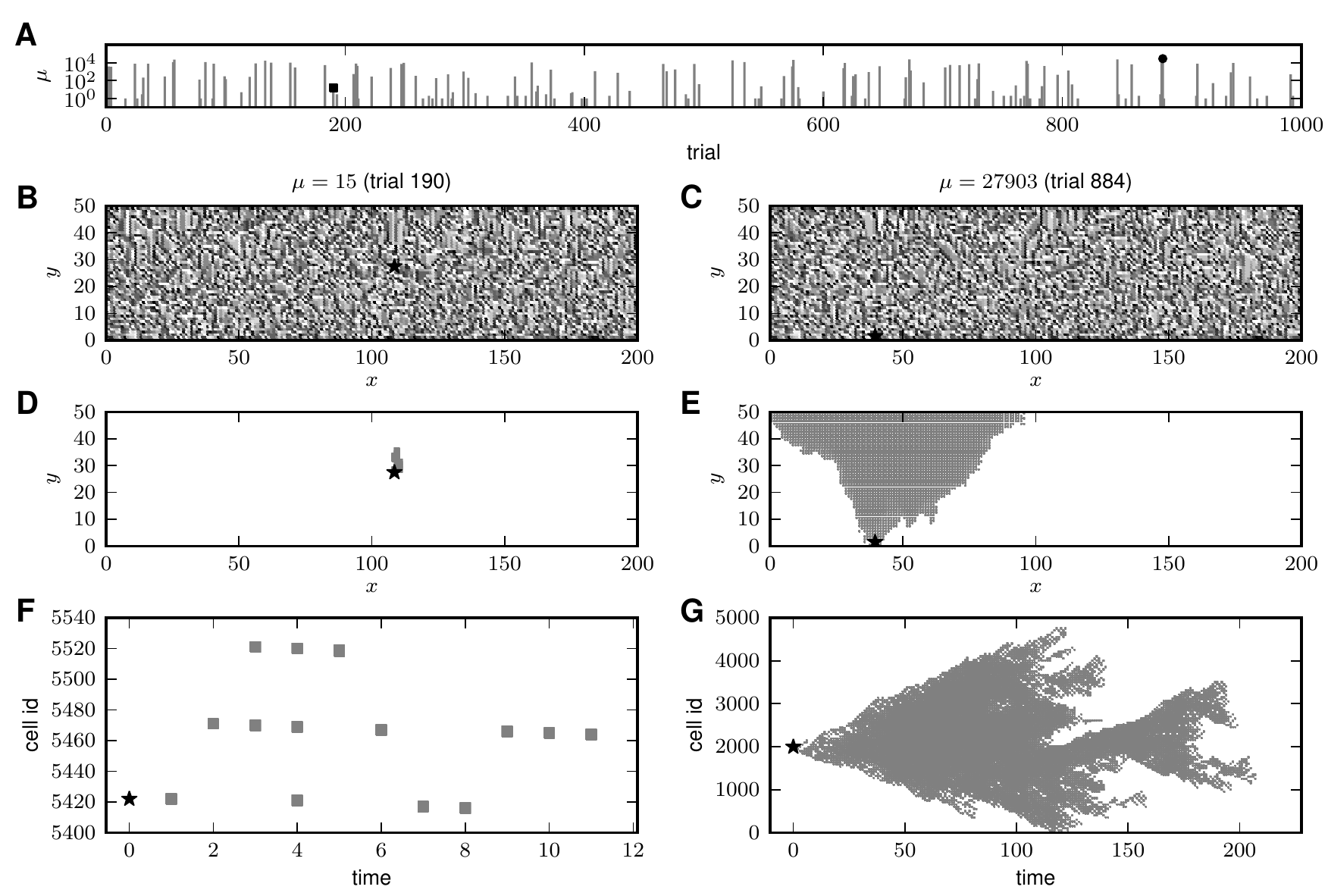}
{
    Size and duration variability of calving events in response to random external perturbations (model results).
    A: Glacier responses (event size $\mu$; log scale) to $1000$ consecutive random perturbations. 
    Black square (trial 190) and circle (trial 884) mark events shown in B,D,F and C,E,G, respectively.
    B--G: Moderate-size (B,D,F; $\mu=15$) and large calving event (C,E,G: $\mu=27903$) in response to punctual 
    random perturbations (perturbation sites indicated by black stars) in trials 190 and 884, respectively.
    B,C: Distribution of stress $z_{xy}$ (gray coded) across the model-glacier terminus after calving 
    response.
    D,E: Spatial spread of single-trial calving activity across the model glacier terminus. Gray dots mark cells which have calved.
    F,G: Spatio-temporal spread of the events shown in D,E (id of cell at position $(x,y)$ $=Hx+y$; 
    discrete time). Note different scales in F and G.
    Glacier width $W=200$, glacier height $H=50$, calving susceptibility $w=1.3$.}
{fig:model_examples}
As in the field data, the distributions of event sizes $\mu$ generated
by the calving model resemble power-law distributions
(\prettyref{fig:model_data}A). The width of the size distribution
increases with the calving susceptibility $w$
(\prettyref{fig:model_data}B) while the power-law exponent
$\gamma_\mu$ decreases and approaches $1$ for $w=1.3$ (solid curve in
\prettyref{fig:model_data}C). Log-likelihood ratios $R$ are always
positive and increase with $w$ (data not shown), thereby indicating
that the power-law model \prettyref{eq:pl_dist} fits the data better
than the exponential model \prettyref{eq:exp_dist}. For $w=1.3$, the
size distribution spans about $6$ orders of magnitudes
($\mu=1,\ldots,10^6$). Note that the maximum event size can exceed the
system size $WH$ (dashed vertical lines in
\prettyref{fig:model_data}A,B) as a single cell can calve several
times during one event (trial).
\par
The inter-event intervals $\tau$ obtained from the calving model span
about $2$ orders of magnitude (\prettyref{fig:model_data}D,E). In
contrast to the event-size distributions, the width of the interval
distribution is independent of the calving susceptibility $w$
(\prettyref{fig:model_data}E). ML fitting of the power-law
\prettyref{eq:pl_dist} and the exponential model
\prettyref{eq:exp_dist} yields exponents $\gamma_\tau$ and
$\lambda_\tau$ which are comparable to those obtained for the field
data (compare \prettyref{fig:model_data}F and
\prettyref{fig:field_data}E--G). The log-likelihood ratios $R$ are
always close to zero (data not shown), again consistent with the field
data.
\par 
Note that the calving variability arising from the model is not a
mere result of the randomness of the external perturbations (see
\prettyref{sec:calving_model}, Experimental protocol). Restricting the
repeated external perturbations to one and same cell in the center
$\{W/2,H/2\}$ of the terminus results in slightly narrower but
qualitatively similar event-size and interval distributions (data not
shown).

\doublecolumnfigure{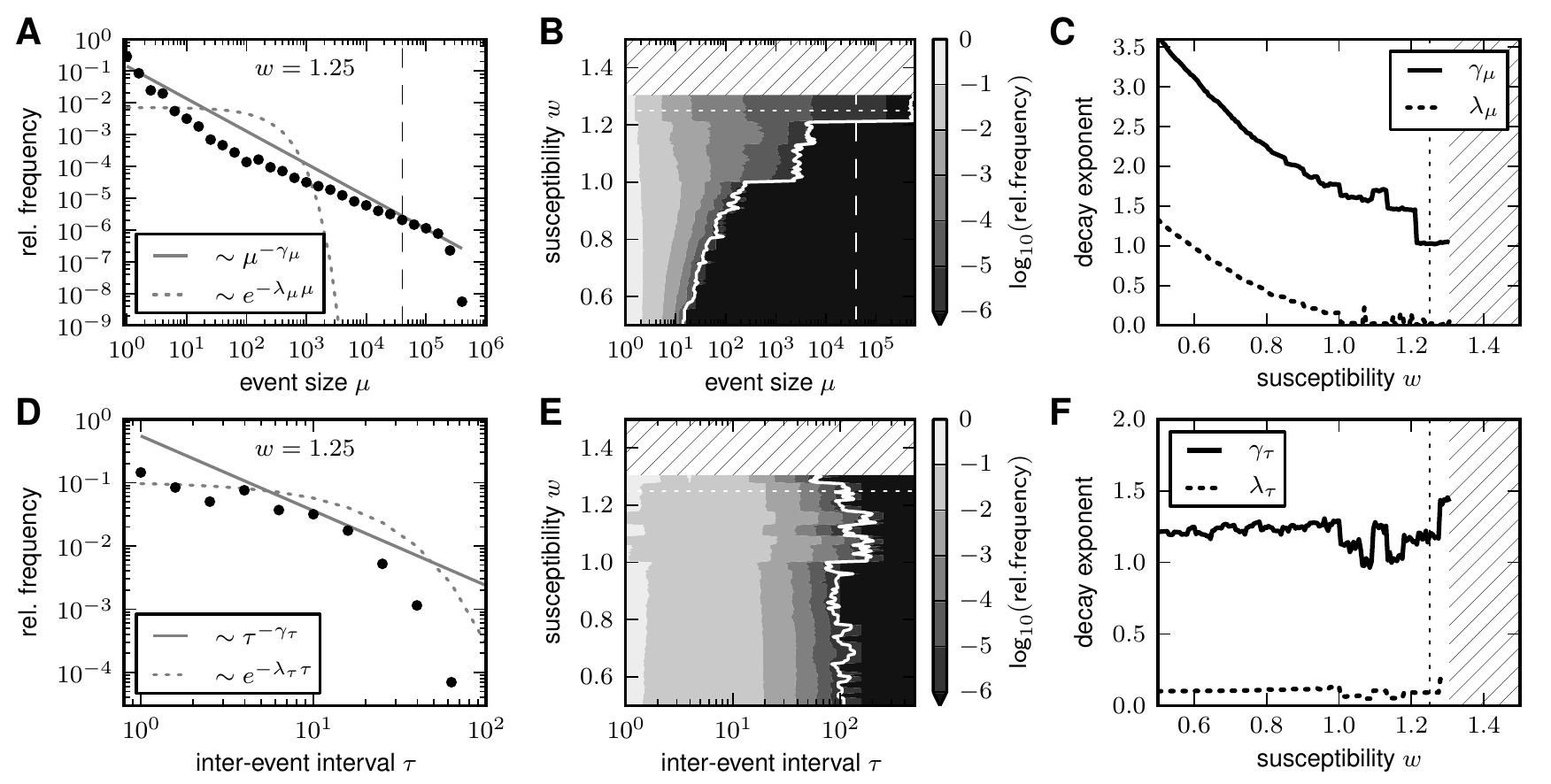}  
{
  Size (top row) and inter-event interval distributions (bottom row) generated by the calving model.
  A,D: Distributions (log-log scale) of iceberg sizes $\mu$ (A) and inter-event intervals $\tau$ 
  (D) for calving susceptibility $w=1.25$. Simulation results (symbols), best-fit power-law (solid lines) 
  and exponential distributions (dashed lines).
  B,E: Dependence of size and interval distributions (log-log scale) on calving susceptibility $w$. 
  Solid white curves mark maximum sizes and inter-event intervals.
  C,F: Dependence of decay exponents of best-fit power-law (solid lines; $\gamma_\mu$, $\gamma_\tau$) 
  and exponential distributions (dotted lines; $\lambda_\mu$, $\lambda_\tau$) on calving susceptibility $w$..
  Glacier width $W=400$, glacier height $H=100$. Vertical dashed lines in A and B indicate system size $WH=40000$.
  Dotted horizontal and vertical lines in B,E and C,F, respectively, mark susceptibility $w=1.25$ used in A and D.
  Hatched areas in B,C,E,F correspond to regions with ongoing, self-sustained calving (see \prettyref{sec:self_sustained}).
}
{fig:model_data}

\subsection{Impact of external parameters}
\label{sec:external_parameters}
As shown in the previous subsection, the calving model generates broad
distributions of event sizes and inter-event intervals, even under
perfectly stationary conditions. Fluctuations in external parameters
are therefore not required to explain the event-size and interval
variability observed in the field data. 
Here, we further support this finding by analyzing the relation
between calving activity and fluctuations in air temperature and tides
during the observation period.
Both tides and temperature can, in principle, affect calving activity.
High tides increase buoyant forces, thereby destabilizing the glacier
terminus. An increase in temperature lowers the yield stress
\citep{Benn2007a} and therefore leads to an increase in the calving
susceptibility $w$. Hence, one may expect that temperature and
tide-level fluctuations lead to changes in the calving-event size and
interval statistics, and thereby explain the large variability
reported in \prettyref{sec:variability} (see
\prettyref{fig:field_data}).
\par
\prettyref{fig:raw_data} depicts the simultaneous time series for
event sizes (A), inter-event intervals (B), air temperature (C),
change in air temperature (D), and tidal amplitude (E) during the
$12$-days observation period in 2009 at Kronebreen. Within this
sampling period, temperatures varied between $-0.8$ and
$8.8\,^\circ\text{C}$, tide levels between $14$ and $178\,\text{cm}$.
Significant correlations between climatic parameters and calving
activity are not observed.
\singlecolumnfigure{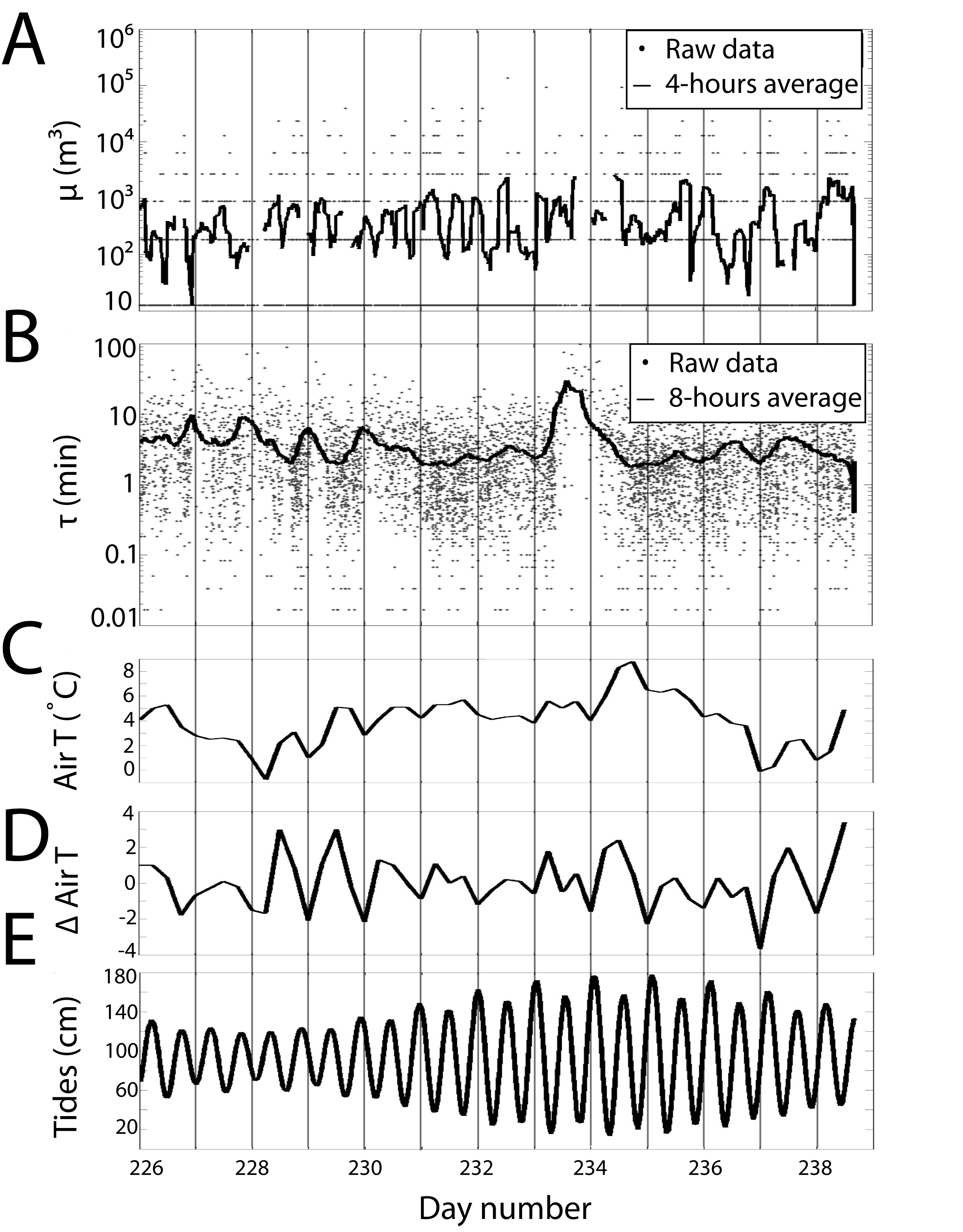}{%
  Relation between calving and external parameters (field data,
  Kronebreen, 2009).
  A: Event sizes $\mu$ (log scale). Individual events (dots) and
  $4$-hour average (black curve).
  B: Inter-event intervals $\tau$ (log scale). Individual events
  (dots) and $8$-hour average (black curve).
  C: Air temperature (Ny-{\AA}lesund, Norwegian Meteorological Institute).
  D: Change in air temperature within $6$ hours.
  E: Tidal amplitude (Norwegian Mapping Authorities).
}  {fig:raw_data}
To test whether fluctuations in air temperature and tidal amplitude
have an effect on the overall shape of the event-size and interval
distribution, we grouped the data into high/low temperature/tide
intervals. The event-size and interval histograms obtained for each
group are indistinguishable (\prettyref{fig:grouped_field_data}).
Hence, the fluctuations in air temperature and tides \emph{within the
  observation period} have no effect on the shape of the
distributions. They cannot explain the observed event-size and
interval variability.
This does not imply that changes in temperature and tides do not
affect the calving statistics in general. Long-term observations are
required to investigate how larger modulations in temperature and tide
levels (e.g.~seasonal fluctuations) affect the shape of the event-size
and interval distributions (see \prettyref{sec:discussion}).
\par
From the results reported here, we cannot exclude the possibility that
fluctuations in other parameters, for example changes in glacier
velocity or buoyancy, contribute to the size and interval
variability. Note, however, that both glacier velocity and buoyancy
are affected by calving itself \citep{Benn2007a}.  They are not purely
``external'' parameters. In our simple model, the positive-feedback
loop between calving and these parameters is abstracted in the form of
the (positive) interaction kernel (see \prettyref{sec:calving_model}).
Hence, correlations between calving activity and fluctuations in
glacier velocity or buoyancy can arise naturally from the local
calving dynamics at the glacier terminus.
\singlecolumnfigure{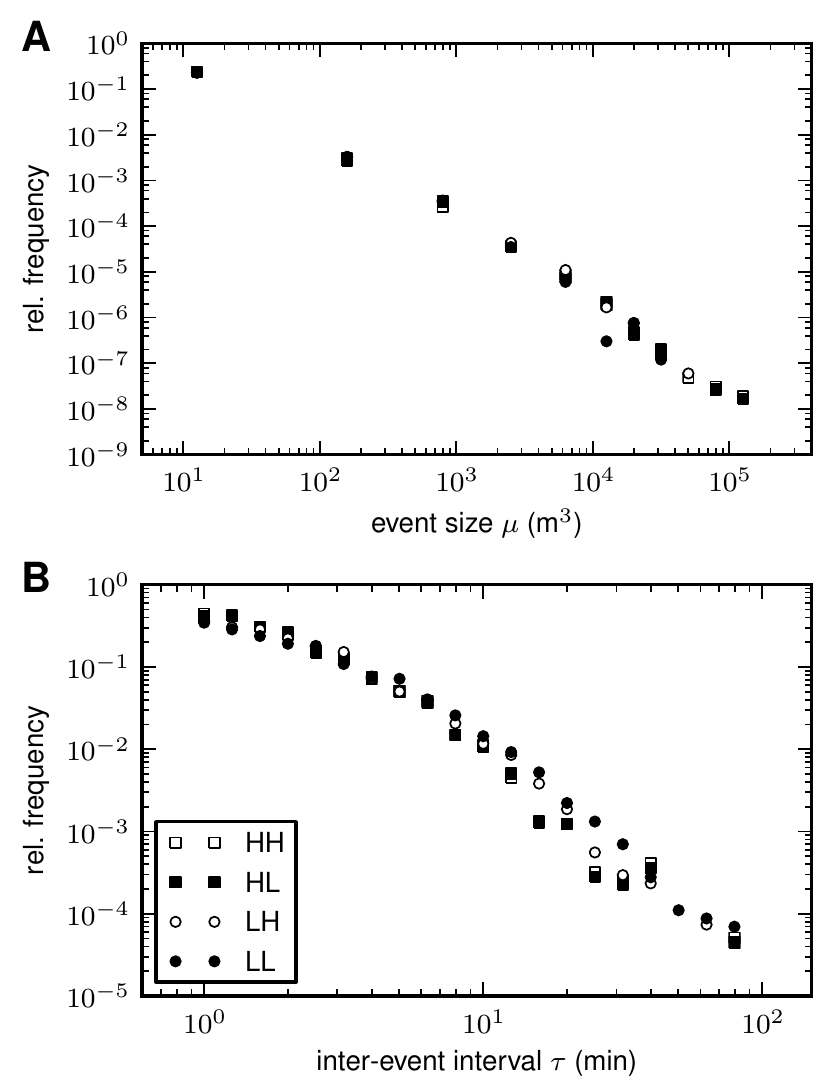}  
{
  Distributions of iceberg sizes $\mu$ (A) and inter-event intervals $\tau$ (B) 
  for four different combinations of air-temperature and tide ranges 
  (Kronebreen, 2009; log-log scaling): 
  high temperature -- high tide (HH),
  high temperature -- low tide (HL),
  low temperature -- high tide (LH),
  low temperature -- low tide (LL).
  Low/High air-temperature interval: $-0.8$--$4.2$/$4.2$--$8.8\,^\circ\text{C}$,
  Low/High tide-level interval: $14$--$92$/$92$--$178\,\text{cm}$.
}
{fig:grouped_field_data}

\subsection{Self-sustained calving}
\label{sec:self_sustained}
For susceptibilities $w\le{}1.3$, small perturbations of the model
glacier result in calving responses with finite life time (see
e.g.~\prettyref{fig:model_examples}F,G). Calving of a single cell can
trigger an avalanche which sooner or later fades away and stops. For
$w>1.3$, the model glacier enters a new regime where the avalanche
life times seem to diverge (hatched areas in
\prettyref{fig:model_data}B,C,E,F). We illustrate this by randomly
initializing the model glacier such that only a small fraction ($1\%$)
of cells is superthreshold ($z>1$) at time $t=0$ and hence starts
calving immediately, simulating the model glacier until this initial
calving stops and measuring the corresponding survival time for a
broad range of susceptibilities $w=0.8,\ldots,1.5$ (see
\prettyref{fig:model_survival}). For $w>1$, the survival time quickly
increases with $w$ and diverges at about $w=1.3$. Beyond this critical
susceptibility, calving never stopped within the maximum simulation
time (here $900$ time steps). Note that this behavior is robust and
does not critically depend on the choice of other model parameters
such as the terminus dimensions ($W$, $H$) and the shape of the
interaction kernel \prettyref{eq:interactions}. In all cases, we
observe a critical susceptibility close to $w=1.3$ (data not shown
here).
\singlecolumnfigure{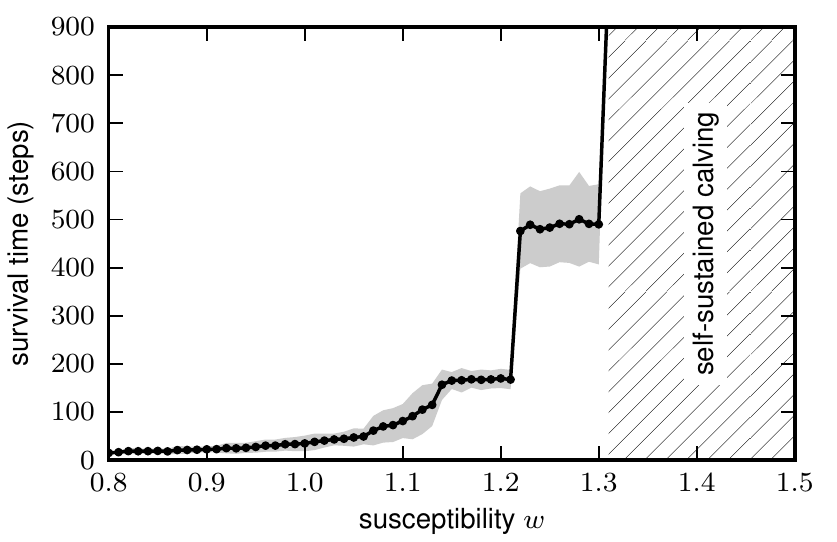}{
  Survival time of calving activity after random stress initialization
  (with $1\%$ of the cells being superthreshold, $z_{xy}>1$, at time
  $t=0$) as function of calving susceptibility $w$. Mean survival time
  (black circles) and mean $\pm 2$ standard deviations (gray band) for
  $100$ random glacier initializations. Glacier width $W=400$, height
  $H=100$.
}
{fig:model_survival}

\subsection{Predictability}
\label{sec:predictability}
In the field data, the average correlation between the sizes $\mu_j$
and $\mu_{j+i}$ of subsequent events $j$ and $j+i$ hardly differs from
zero for all $i>0$ (\prettyref{fig:correlations}A). Hence, predicting
future event sizes from past events appears hopeless. This behavior
is reproduced by the simple calving model
(\prettyref{fig:correlations}B).
The inter-event intervals $\tau$ obtained from the field data exhibit
a moderate long lasting correlation which is not observed in the model
data (\prettyref{fig:correlations}C,D). The reason of this discrepancy
between the model and the field data is unclear, but we suspect that
it is due to non-stationarities in the field data (see
\prettyref{fig:raw_data}B) which are, by construction, absent in the
model.
\singlecolumnfigure{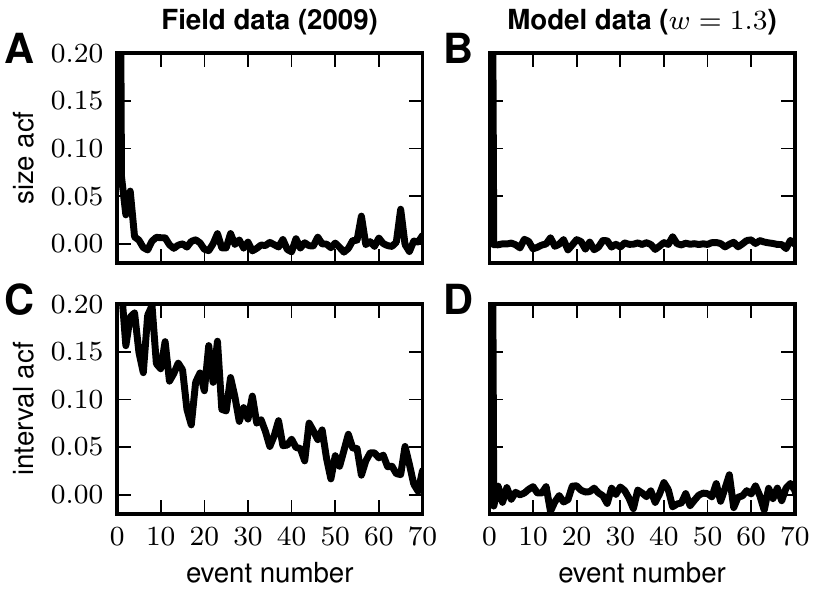}{
  Auto-correlation function (acf) for sizes (A, B) and inter-event
  intervals (C,D) in the field (A,C) and model data (B,D).
}
{fig:correlations}

\section{Discussion}
\label{sec:discussion}
We showed that calving-event sizes and inter-event intervals obtained
from direct continuous observations at the termini of two tidewater
glaciers are highly variable and broadly distributed. We demonstrated
that the observed variability can be fully explained by the mutual
interplay between calving and the destabilization of the local
neighborhood of the calving region. A simple calving model accounting
for this interplay reproduces the main characteristics of the
event-size and interval statistics: i) Event-size distributions
resemble power laws with long tails spanning several orders of
magnitude. ii) Interval distributions are broad but fall off more
rapidly than power laws. iii) Correlations between the sizes of
subsequent events vanish. We conclude that the observed calving
variability is a characteristic feature of calving and is not
primarily the result of fluctuating external (e.g.~climatic)
conditions. Event sizes of all magnitudes have to be expected, even
under ideal stationary conditions.
\par
The calving model predicts that the width of the event-size
distribution depends on a parameter $w$ representing the calving
susceptibility of the glacier terminus: roughly, $w$ measures how
prone the glacier is to calve in response to calving. It describes to
what extent calving increases the internal ice stress in the
neighborhood of the calving region. Alternatively, $w$ may represent
the inverse of the yield stress, i.e.~the critical stress at which ice
breaks. The susceptibility $w$ is determined by the properties of the
ice and by factors like temperature, glacier velocity, buoyancy or
glacier thickness. In our model, the width of the event-size
distribution increases monotonously with $w$. Therefore, the shape of
the size distribution, as, for example, characterized by the power-law
exponent $\gamma_\mu$, may be informative about the stability of the
glacier terminus. In contrast to the event-size distributions, the
shape of the inter-event interval distribution is insensitive to the
susceptibility $w$. At a critical susceptibility $w_\text{crit}$, the
model predicts an abrupt transition of the glacier to a new regime
characterized by ongoing, self-sustained calving activity.
Observations of rapid glacier retreats \citep{Pfeffer2007, Briner2009,
  Motyka2011} may be explained by these supercritical dynamics
\citep[see also][]{Amundson2010}.
In the model, the power-law exponent $\gamma_\mu$ is close to $1$ as
the susceptibility approaches the critical value $w_\text{crit}$ (see
solid curve in \prettyref{fig:model_data}C). We found that this
observation does not critically depend on the choice of the model
parameters (glacier dimensions, shape of the interaction kernel,
perturbation protocol). This suggests that the power-law exponent
$\gamma_\mu$ may serve as an indicator of a glacier's proximity to the
transition point where it starts retreating
rapidly. \citet{Bassis2010} proposed a similar stability criterion
which depends on various geometric and dynamical near-terminus
parameters.  In our case, the diagnostics is exclusively based on the
distribution of event sizes.
\par
The calving model in this study is inspired by previous work on the
emergence of power-law distributions. Power-law shaped magnitude
distributions are abundant in nature. They are found, for example, for
earthquake magnitudes \citep{Bak2002, Hainzl2003}, luminosity of
stars \citep{Bak96}, avalanche sizes in sandpiles \citep{Bak96},
landslide areas \citep{Guzzetti2002}, subglacial water pressure pulses
\citep{Kavanaugh2009}, dislocation avalanches in ice
\citep{Richeton2005}, and sea ice fracturing
\citep{Rampal2008}. \citet{Bak87} demonstrated that power-law
distributions can arise naturally in spatially extended dynamical
systems which have evolved into ``self-organized critical states''
consisting of minimally stable clusters of all length scales.
Perturbations can propagate through the system and evoke responses
characterized by the absence of spatial and temporal scales
(avalanches). The calving model used in the present study is
qualitatively similar to the sandpile model of \citet{Bak87}. It is
therefore not surprising that it predicts power-law shaped event-size
distributions. The scientific value of this part of our work exists in
mapping the original model by \citet{Bak87} to the dynamics of glacier
calving and in relating the model parameters to physical
measures. Moreover, the phenomenon of self-sustained ongoing activity
has, to our knowledge, never been reported so far in this context.
\par
Several aspects of this work may be subject to criticism and
improvement, in particular the data acquisition and the construction
of the calving model:
direct visual observation by humans is so far one of the most reliable
methods to monitor individual calving events in continuous time
\citep{vanderveen97}. However, the results are hard to reproduce
exactly: due to the sporadic unpredictable nature of calving, it is
difficult to capture each event. Different observers may be in
different attentive states. Each new observer needs to adjust her- or
himself to a common scale of perceived sizes $\psi$. To minimize the
variability in perceived sizes, we arranged test observations where all
observers were simultaneously confronted with size estimates (see
\prettyref{sec:field_data_acquisition}).
To confirm our findings on the relation between the shape of the
event-size distribution and the state of the glacier, more data are
needed from glaciers in different environments (freshwater, floating
tongue, iceshelves) and different dynamical states (advancing, stable,
retreating). To test our stability criterion one would need to
continuously monitor the same glacier year after year for a few
days. Long-term continuous observations or observations in cold months
are difficult.
Automatic monitoring would therefore be highly
desirable. Unfortunately, all available automatic monitoring methods
have limitations. Terrestrial photogrammetry, for example, is limited
by the iceberg size, visibility and illumination of the
glacier. Iceberg size and type also limit the use of ground-based
radar. \citet{Chapuis2010} showed that radar could only detect events
larger than $150\,\meter^3$. Remote sensing (optical and radar
imagery) has the same limitations as terrestrial photogrammetry. In
addition, its low temporal resolution does not allow individual
calving events to be registered. Seismic monitoring \citep{ONeel2010}
is a very promising technique, but can detect only the largest events
\citep{Koehler12_393}. The range of event sizes accessible by seismic
methods may however be sufficient to estimate power-law exponents.
A major problem in studying the statistics of \emph{individual}
calving events is to define what a single event actually is. In the
framework of our model, an event is defined as the total response to a
single perturbation. In our numerical experiments, a new perturbation
is not applied before the response to the previous perturbation has
stopped. In nature, however, the glacier terminus may be constantly
perturbed, e.g.~by the movement of the glacier. Several events may be
triggered simultaneously in neighboring regions of the terminus and,
hence, overlap both spatially and temporally. The separation of
individual events in the field data can therefore be difficult. In
consequence, short intervals and small events may be underestimated.
\par
The calving model used in this study is highly minimalistic. It
implements the mechanism which we think is essential for an
understanding of the calving variability observed in the field data
(i.e.~a positive-feedback loop between calving and destabilization of
the glacier terminus), but neglects a variety of factors:
first, the model is two-dimensional. Stress can only propagate
tangentially and not in the third dimension perpendicular to the
terminus. Extending the model to three dimensions is not straightforward,
as it is unclear how calving at the terminus affects regions deeper
within the ice body.
Second, the interaction kernel is restricted to nearest-neighbor
interactions. Whether and how calving affects regions more distant
from the calving region is unclear. A direct measurement of the
interaction kernel in the field is difficult as it would require
monitoring of changes in ice stress in response to individual calving
events. Conclusions on the shape of the interaction kernel could be
drawn indirectly from the spatial structure of calving avalanches
(e.g.~obtained by terrestrial photogrammetry). For illustration,
consider \prettyref{fig:model_examples}E: the triangular shape of this
large event is a direct consequence of the asymmetry of the
interaction kernel \prettyref{eq:interactions_asymmetric} (see
\prettyref{fig:model_sketch}C). For symmetric kernels
\prettyref{eq:interactions_symmetric}, we observed that the spatial
structure of calving events is, on average, symmetric (data not shown
here). Note, however, that for the main findings of our study, the
exact shape of the interaction kernel is not critical.
Third, the relationships between the susceptibility $w$ (the area of
the interaction kernel, see \prettyref{sec:calving_model}) and
external parameters such as air temperature, tides, glacier velocity
or buoyancy are unclear. These relationships could be established
empirically from a larger data set obtained from continuous long-term
observations of different glaciers during different dynamical
states. For each identified set of external parameters, our model
could be fitted to the distribution of monitored event sizes, thereby
providing an estimate of $w$. Based on these relationships, the
event-size distributions for a new ``test'' data set (i.e.~for data
not used for the fitting procedure) could be predicted by the model
and compared to the field-data distributions.
Fourth, stress increments in response to calving or external
perturbations are instantaneous in the model. In reality, these
interactions are likely to be smoother and delayed. Detailed
information about this is however limited.
Fifth, in the absence of calving and external perturbations,
internal ice stress in a given model cell is constant. In real ice,
stress in a certain volume element may slowly dissipate and decay to
zero. We assume that the time constant of this decay is much larger
than the time between two (external or internal) perturbations and
therefore can be approximated as being infinite. 
Sixth, external perturbations are modeled as punctual (delta-shaped)
events in time and space to characterize the system's response in a
well defined manner (i.e.~to evoke responses without spatio-temporal
overlap; see above). In reality, external perturbations are spatially
and temporally distributed (e.g.~glacier movement, glacier velocity
gradients, buoyancy).
Finally, the model neglects submarine calving which, in reality,
amounts for about $13\%$ of the total ice loss at the terminus. One way
to account for the submarine dynamics would be to assign different
calving susceptibilities to the subaerial and the submarine parts.
Our model describes calving at the level of individual events and
focuses on a single aspect of calving: the interplay between calving
and terminus destabilization. Factors like air temperature, water depth,
height-above-buoyancy, surface melt, ice thickness, thickness
gradient, strain rate, mass-balance rate, backward melting of the
terminus, etc., are included indirectly or lumped together and treated
as constant parameters. Previous macroscopic models
\citep{Brown82,Vanderveen96,Vieli2001,Vieli2002,Benn2007b,Amundson2010,Nick2010,
  Otero2010,Oerlemans2011} relate these factors to the overall calving
rate. A description of size and timing of individual calving events is
beyond the scope of most of these models. Our model could be
connected to these models by linking the overall calving rate to the
calving susceptibility or the characteristics of external
perturbations. The resulting multi-scale model would have the
potential to describe the statistics of event sizes and inter-event
intervals under more realistic non-stationary conditions.
The rationale behind the simplistic approach presented in our study is
the opposite: to construct a minimal model which reproduces the main
characteristics of the event-size and interval
statistics. Disconnecting the glacier terminus from the external world
allows us to study calving dynamics under ideal stationary conditions
and to demonstrate that the observed calving variability is not
primarily a result of fluctuations in external conditions, but can
emerge from simple principles governing the dynamics directly at the
glacier terminus.

\section{Acknowledgments}
\label{sec:thanks}
We acknowledge support by the Research Council of Norway
(IPY-GLACIODYN [project 176076], eVITA [eNEURO], Notur), the Helmholtz
Association (HASB and portfolio theme SMHB), the J\"ulich Aachen
Research Alliance (JARA), EU Grant 269921 (BrainScaleS) and the
Svalbard Science Forum. We thank the ``Young Explorers and Leaders''
of the ``British Schools Exploring Society Arctic Adventure''
expedition in 2010 for providing the iceberg-calving data for
Sveabreen. We are grateful to Abigail Morrison, Cecilie Denby-Rolstad
and the two reviewers for their valuable comments on the manuscript.



\end{document}